\documentclass[aps,prb,twocolumn,superscriptaddress,longbibliography]{revtex4-2}
\usepackage[pdftex, pdftitle={Collinear spin density wave state in distorted square-lattice GdNiSn$_4$}, colorlinks,allcolors=blue]{hyperref}
\usepackage{graphicx}
\usepackage{booktabs}
\usepackage{cleveref}
\begin{document}

\title{Collinear spin density wave state in distorted square-lattice GdNiSn$_4$}

\author{Charles C. Tam}
\affiliation{Materials Department, University of California Santa Barbara, Santa Barbara 93106 USA}

\author{Sarah Schwarz}
\affiliation{Materials Department, University of California Santa Barbara, Santa Barbara 93106 USA}

\author{Xin Zhang}
\affiliation{Department of Chemistry, Princeton University, Princeton, New Jersey 08544, USA}

\author{Sudipta Chatterjee}
\affiliation{Department of Chemistry, Princeton University, Princeton, New Jersey 08544, USA}

\author{Scott B. Lee}
\affiliation{Department of Chemistry, Princeton University, Princeton, New Jersey 08544, USA}

\author{Rebecca Scatena}
\affiliation{Diamond Light Source, Harwell Science and Innovation Campus, Didcot, Oxfordshire OX11 0DE, UK}



\author{Leslie M. Schoop}
\affiliation{Department of Chemistry, Princeton University, Princeton, New Jersey 08544, USA}

\author{Stephen D. Wilson}
\affiliation{Materials Department, University of California Santa Barbara, Santa Barbara 93106 USA}


\begin{abstract}

We characterize the magnetic ground state of the newly synthesized lanthanide intermetallic GdNiSn$_4$ via resonant elastic x-ray scattering measurements. This compound forms distorted square nets of Gd that initially order magnetically below 23 K followed by a lower temperature transition at 16 K.  Our scattering data identify the ground state order as a single-\textbf{q} incommensurate, collinear order that slides towards a commensurate wave vector above the 16 K transition. Magnetic symmetry analysis combined with azimuthal dependence resolves the ground state magnetic structure as a moment-modulated spin density wave state with Gd moments oriented parallel to the in-plane $a$-axis. We discuss connections between the observed magnetic order and electronic properties in this square-net compound.

\end{abstract}

\maketitle

\section{Introduction}

One strategy of engineering quantum materials that host unconventional magnetotransport phenomena is to embed a high symmetry network of local lanthanide moments, such as Gd, into crystal lattices that host high mobility (often topologically nontrivial) electronic band structures. The localized lanthanide moments hybridize and interact with the mobile conduction electrons and can often be described by a Hamiltonian with local Kondo and extended Ruderman-Kittel-Kasuya-Yosida (RKKY) interactions. A variety of noncoplanar spin textures can result and generate a finite scalar spin chirality, promoting a topological Hall response. The most widely studied manifestation of this occurs in non-centrosymmetric systems, where a global Dzyaloshinskii–Moriya interaction (DMI) is allowed and skyrmion topological spin textures may form. Common examples are found in B20 alloys such as MnSi and FeGe ~\cite{Muehlbauer2009,Yu2010}, though a much broader range of compounds are known to exhibit similar phenomenology \cite{tokura2020magnetic}. 

For the broader class of centrosymmetric lattices, noncoplanar spin textures and skyrmion phases may also form via RKKY interactions, even in the absence of a global DMI~\cite{Ozawa2017,Hayami2017,Wang2020}. To date, various types of centrosymmetric lattices have been shown to host topological spin textures, particularly among geometrically frustrated lattices such as triangular Gd$_2$PdSi$_3$~\cite{Kurumaji2019,Hirschberger2020} and kagome Gd$_3$Ru$_4$Al$_{12}$~\cite{Hirschberger2019} with multi-\textbf{q} order. More recently, similar phenomenology was discovered in tetragonal lattices that are free from conventional geometric frustration, with examples including EuAl$_4$~\cite{Nakamura2015,Kaneko2021,Takagi2022}, EuGa$_2$Al$_2$~\cite{Moya2022},  GdRu$_2$Si$_2$~\cite{Garnier1995,Yasui2020,Khanh2020,Khanh2022}, and GdRu$_2$Ge$_2$~\cite{Garnier1996,Yoshimochi2024}. This opens a door for an even larger pool of non-hexagonal, centrosymmetric compounds and motivates the use of square-nets of lanthanides as a building block for engineering topological spin textures into high-mobility intermetallic compounds as a means of creating unconventional Hall responses. 

\begin{figure}[t]
\includegraphics[]{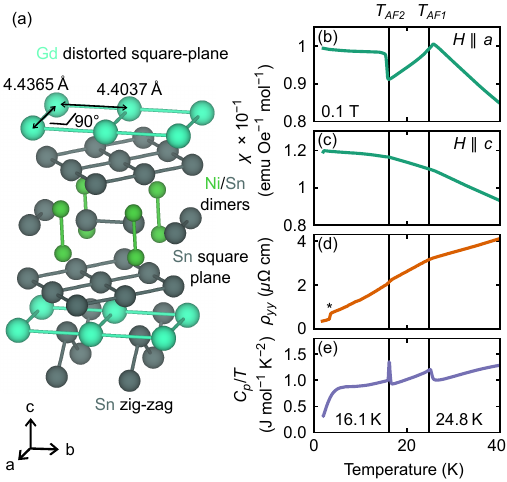}
\caption{\textbf{Physical properties of distorted square-lattice GdNiSn$_4$.} (a) Crystal lattice of GdNiSn$_4$ where differing structural motifs are labeled. The structure has been simplified for clarity; more detailed structural information is given in Ref.~\onlinecite{Zhang2025}. Temperature dependent magnetic susceptibility with the applied field of $\mu_0 H = 0.1\;$T is shown applied (b) parallel to the $a$-axis, (c) and parallel to the $c$-axis. (d) Zero field $\rho_{yy}$. Note the transition at $\approx 4\;$K (denoted by $*$), is a a superconducting transition of residual Sn flux. (e) Specific heat capacity divided by temperature $C_p/T$ collected at zero field.}
\label{fig:fig1} 
\end{figure}

In a recent report, Zhang \textit{et al.} synthesized and characterized a new intermetallic compound GdNiSn$_4$~\cite{Zhang2025}. This compound is monoclinic and forms bilayers of distorted square-planes of Gd. They are slightly distorted square nets, because although the angle between $a$ and $b$ is $90^\circ$, the nearest neighbor Gd-Gd bond distance is inequivalent, with Gd-Gd along $a$ being 4.4365\;$\mathrm{\AA}$ and 4.4037\;$\mathrm{\AA}$ along $b$ (see \Cref{fig:fig1}a). The nature of the ground state magnetic order and its evolution under magnetic field remain unknown.

In this paper, we determine the ground state magnetic structure of GdNiSn$_4$ and its evolution upon warming through $T_{SDW}$.  Due to strong nuclear absorption associated with Gd ions, resonant elastic x-ray scattering (REXS) techniques were utilized rather than traditional neutron diffraction. Single-\textbf{q}, incommensurate magnetic order is observed with moments modulated both within the Gd square-net $ab$-plane as well as along the interlayer direction. Magnetic symmetry analysis combined with azimuthal dependence restricts the magnetic structure to a collinear spin density wave with moments parallel to the $a$-axis.  The coupling of this magnetic structure to the electronic properties of GdNiSn$_4$ is discussed.



\section{Methods}

GdNiSn$_4$ single crystals were grown via a Sn-based self-flux technique. More details on synthesis and characterization is given in Ref.~\onlinecite{Zhang2025}. GdNiSn$_4$ crystallizes in the $C/2m$ space group, with two distinct Gd atoms on the $4i$ Wykoff site. Here we label the $c$-axis as the long axis, and the $ab$-plane is the basal plane in which the Gd square nets reside.

Resistivity and specific heat data were collected using a Quantum Design Physical Property Measurement System (Dynacool PPMS). Resistivity measurements were performed a four-probe method with contacts created with silver paste, and data were collected using the electrical transport option of the PPMS. Specific heat measurements were performed using the quasi-adiabatic relaxation method and the crystal mounted within the heat capacity option of the PPMS. Orientation-dependent magnetization data were collected using a Quantum Design Magnetic Property Measurement System (MPMS3).

Resonant elastic x-ray scattering (REXS) measurements were performed at beamline I16, Diamond light source~\cite{Collins2010}. Samples were measured as as-grown and mounted in a reflection geometry, using a beam size of $\sim 180 \mu\mathrm{m} \times 25 \mu\mathrm{m} $ (width $\times$ height). Specifically, samples were mounted such that the (001) was the specular direction, and (100) was the azimuthal reference. Samples were cooled with an ARS closed-cycle cryostat with a beryllium dome. Momentum transfer is labeled in reciprocal lattice units (r.l.u.) of the monoclinic unit cell of GdNiSn$_4$, where $\mathbf{Q} = H \mathbf{a}^{\star} + K \mathbf{b}^{\star} + L \mathbf{c}^{\star}$, $a = 8.7409\;\mathrm{\AA}$, $b = 8.8238\;\mathrm{\AA}$, $c = 14.3341\;\mathrm{\AA}$, and $\alpha = \gamma = 90^\circ$, $\beta = 98.77^\circ$. In a vertical scatting geometry, incident light from the undulator is horizontally polarized, and is therefore $\sigma$ polarized with respect to the scattering plane. Linear polarization analysis was done using a graphite (006) crystal oriented close 44.4$^\circ$, and the incident polarization was controlled using light transmitted through a $400\mu\mathrm{m}$ diamond (001) plate oriented close to the (111) reflection in asymmetric Laue mode.

\section{Experimental Results}

Measurements of the temperature dependent DC magnetic susceptibility of GdNiSn$_4$ are plotted in \Cref{fig:fig1}(b,c). Consistent with measurements reported in \cite{Zhang2025}, the data indicate antiferromagnetic order appears initially at $T_{AF1}\approx 25$~K with a second transition near $T_{AF2}\approx 16$~K. Orientation dependent measurements suggest an enhanced susceptibility within the $ab$-plane, and the two transitions are consistent with anomalies within the resistivity $\rho_{yy}$ which is plotted in \Cref{fig:fig1}d., and zero-field specific heat, which is plotted in \Cref{fig:fig1}e.

\begin{figure}[h]
\includegraphics[]{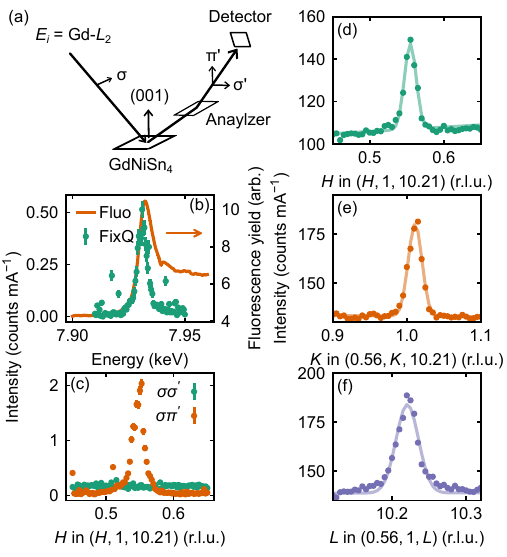}
\caption{\textbf{Incommensurate magnetic order in GdNiSn$_4$.} (a) Scattering geometry and linear polarization analysis schematic. (b) Energy dependence of the (0.56, 1, 10.22) peak, and fluorescence on the secondary axis. (c) $H$ dependence of magnetic peak with $\sigma \sigma'$ and $\sigma \pi'$ incoming and outgoing polarization. (d)-(f) High symmetry cuts of the magnetic peak along $H$, $K$ and $L$. All measurement were taken at 7.5\;K.}
\label{fig:fig2} 
\end{figure}

In order to characterize the magnetic structure responsible for the anomalies in C$_P$(T), resonant x-ray diffraction measurements were performed.  The incident x-ray energy was tuned to the Gd-$L_2$ absorption peak near 7.933~keV, enhancing resonant scattering coming from the Gd atoms.
After a thorough search through reciprocal space, an incommensurate peak was observed at $\mathbf{Q} = (0.56, 1, 10.22)$ and equivalent positions, demonstrating an incommensurate magnetic \textbf{q}=(0.06, 0, 0.22). In \Cref{fig:fig1}c, a fixed-\textbf{Q} energy scan on a magnetic Bragg reflection renders a distinct resonant profile, peaking slightly below Gd-$L_2$ edge ($\approx 2\;$eV). Magnetic scattering is expected to rotate linearly polarized light by $90^\circ$, and in \Cref{fig:fig2}c, polarization analysis of the scattered beam shows this to be the case. In a broader search of symmetry equivalent positions in different Brillouin zones, these peaks are only present at odd $H$ (\Cref{fig:fig2}d) and odd $K$ (\Cref{fig:fig2}e) zone centers while they exist for both even and odd $L$ (\Cref{fig:fig1}f). 

\begin{figure}[t]
\includegraphics[]{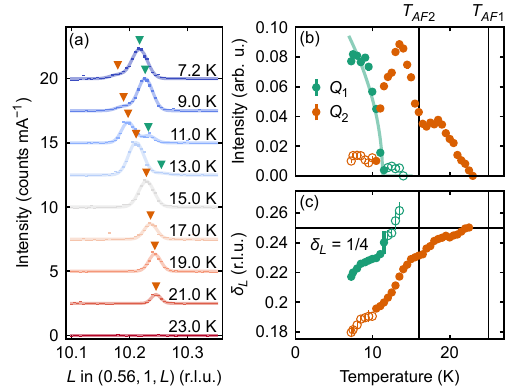}
\caption{\textbf{Temperature dependence of magnetic order of GdNiSn$_4$.} $L$ scans taken at different temperatures across the (0.56, 1, 10.22) peak. The background has been subtracted and higher temperature scans have been offset for clarity. Solid lines are fits. Peak centers are marked. (b) Temperature dependence of the two fitted peak intensities. The solid green line as an order parameter fit to the $Q_1$ intensity, and the solid orange line is a Gaussian fit to the $Q_2$ intensity. (c) Temperature dependence of the two fitted peak centers. Hollow markers in (b) and (c) are the fitted values of the peak when intensity is too low compared to the other peak to be reliable.}
\label{fig:fig3} 
\end{figure}

The temperature dependence of the magnetic peak was measured and is plotted in \Cref{fig:fig3}a. At 7.5 K, it becomes clear the magnetic peak is composed of a primary peak at $Q_1 = (0.56, 1, 10.22)$ with a secondary peak visible as a shoulder at $Q_2 = (0.56, 1, 10.18)$. Upon warming, scattering at $Q_1$ is suppressed with an order parameter-like behavior and is no longer detectable above $T_{AF2}=13$~K. Meanwhile, upon warming the scattering along $Q_2$ grows in intensity and is peaked near 13~K. Warming up further, $Q_2$ is then subsequently suppressed until it is no longer detected around $T_{AF1}=23$~K. This behavior is consistent when measured across equivalent positions in different zones (see supplementary information \cite{Supplemental}). We note there is an apparent $\approx$3~K offset in the x-ray-derived transition temperatures relative to the $T_{AF1}$ and $T_{AF2}$ transitions identified from magnetization, heat capacity, and resistivity data in Figure 1.  This apparent shift downward in the onset temperatures of transitions observed via x-ray scattering likely arises from local beam heating of the sample by the x-ray source.

Now turning to the incommensurability $\delta$ of the peaks (i.e. $\delta_L$, where $L = 10 + \delta_L$) as a function of temperature in \Cref{fig:fig3}c, both peaks at $Q_1$ and $Q_2$ show continuous changes when warmed; however the rate of change of $Q_2$ with temperature slows considerably when order along $Q_1$ vanishes and peak trends to a commensurate value of $\delta=1/4$ as $T_{AF2}$ is approached. Extrapolating linearly to $T=0$ K, $Q_1$ projects to a ground state incommensurability of $\delta \approx 0.18$, while the coexisting $Q_2$ harmonic becomes strongly suppressed.  The in-plane wave vector remains fixed upon cooling through both phases at $\delta_H=0.06$



\begin{figure*}[t!]
\includegraphics[]{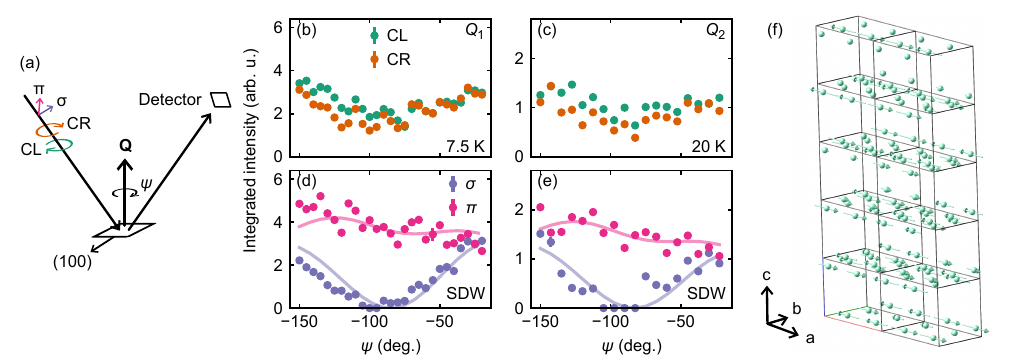}
\caption{\textbf{Determination of zero-field magnetic structure of GdNiSn$_4$.} (a) Scattering geometry schematic for azimuthal dependent scans with varying incident polarization and without polarization analysis. (b), (c) Integrated intensities of rocking curves on $Q_1 = (0.56, 1, 10.22)$ at 7.5~K and $Q_2 = (0.56, 1, 10.24)$ at 20~K, with circularly polarized incident light. (d) ,(e) Integrated intensities of rocking curves on $Q_1$ at 7.5~K and $Q_2$ at 20~K with linearly polarized light. Solid lines are fits to the proposed magnetic structure, with a best fit value ofhe  cant angle $\phi_1 = 99.9^\circ\pm1.5^\circ$ at 7.5~K, and  $\phi_2 = 103.5^\circ\pm2.1^\circ$ at 20~K. $\phi$ is the angle of the moments from the $c$ axis. (f) Visualization of moment-modulated spin density wave structure, where $m \parallel a$. The $c$ axis has been shrunk for clarity.}
\label{fig:fig4} 
\end{figure*}

REXS measurements further allow for an unambiguous measurement of the moment direction in the ordered state. This is accomplished by measuring the azimuthal dependence of the magnetic peak with different incident light polarization. At each azimuth value in the scan, a rocking curve is performed and the intensity integrated for the magnetic peak. The azimuthal value where the intensity is maximized is, to first order, the moment direction. A schematic of this is drawn in \Cref{fig:fig4}a. An azimuthal measurement following this protocol was performed at 7.5~K to probe the moment orientation at $Q_1=(0.56,1,10.22)$, and another at 20~K to ensure $Q_1$ was fully suppressed to probe the moment orientation at $Q_2=(0.56,1,10.24)$. 

The azimuthal dependence was first measured with circularly polarized light, with circular left (CL) and circular right (CR), polarizations. This is plotted in \Cref{fig:fig4}b,d for measurements taken at 7.5~K and 20~K respectively. The absence of contrast between CL and CR is consistent with a collinear magnetic structure. With linearly polarized $\sigma$ and $\pi$ light, there is a total suppression of intensity in the $\sigma$ polarized light at around $-90^\circ$. Since the azimuthal reference is along (100), this means that the maximum in the $\sigma$ polarized channel is near $\phi = 0^\circ$, and therefore that the moments are in the $(a,c)$ plane. Given that the structure is collinear and the ordering wave vector is incommensurate, magnetic symmetry analysis dictates that the structure must be a moment modulated spin density wave (SDW)~\cite{Supplemental}.

The cant direction of the SDW can be determined by fitting the azimuthal dependence to simulated x-ray scattering intensities. Using expressions from Hill and McMorrow~\cite{Hill1996} and taking into account absorption due to resonance, the azimuthal dependence of the data was fit to a model where moments are constrained to the $(a,c)$ plane (where $\phi$ is the cant angle from the $c$ axis). The result is plotted in \Cref{fig:fig4}c where least squares fitting of this model to the data determines $\phi_1 = 99.9^\circ\pm1.5^\circ$ at 7.5~K, and $\phi_2 = 103.5^\circ\pm2.1^\circ$ at 20~K. This is within error of the monoclinic angle $\beta = 98.77^\circ$. Within error of the technique, the moment is parallel to $a$, and the lack of moment projected along $c$ is consistent with the almost featureless $\chi(T)$ when field is applied along the $c$ axis (\Cref{fig:fig1}d). A representative picture of the resulting collinear magnetic structure in the $mB_2$ irrep is depicted in \Cref{fig:fig4}d. We note that with our current REXS data, we are unable to determine the magnitude of the ordered moment that is modulated.

\section{Discussion and Conclusions}


Sliding of magnetic ordering wave vectors often occurs in spin density wave systems and is reported in other Gd intermetallics~\cite{Porter2023}. A common tendency is for the wave vector to transition or slide toward a commensurate position as a means of minimizing entropy upon cooling. However, this is distinct from the case of GdNiSn$_4$ where the interplanar modulation moves away from commensuration upon cooling. This is likely due to the presence of an intermediate, first-order transition at $T_{AF2}\approx$ 13 K.


There is an extended regime of coexistence between the $T_{AF1}$ and $T_{AF2}$ phases, indicating a first-order phase boundary and consistent with the sharp $C_P$($T$) anomaly shown in Figure 1 (e). This precludes the two $Q$s from corresponding to a single phase multi-$Q$ structure (there are no peaks at $Q_1 + Q_2$), and instead indicates phase separation as the sample is cooled across $T_{AF1}$. The tradeoff in intensities between the two phases is consistent with a metastable regime where the modulation associated with $T_{AF2}$ persists and a different harmonic appears closer to the commensurate point before the $T_{AF2}$ order vanishes. We can also preclude that these two different $Q$s are associated with different crystallographic grains/twins. The twinning of the crystal we measured was captured with a twin law model, which gave a twinning ratio of 57.1/42.9~\cite{Supplemental}. Although this is fairly even, and $Q_1$ and $Q_2$ have fairy consistent peak intensities, their temperature dependence's are distinct. Additionally, the small beam size allowed us to raster on the sample to a spot where no twinning was observed in the structural Bragg peaks in the areas probed by the x-ray beam.  Additional details on the twinning model can be seen in the supplementary information \cite{Supplemental} and Ref.~\onlinecite{Zhang2025}.

The two $T_{AF1}$ and $T_{AF2}$ states only differ in their modulation along the interplane direction and scattering from both states is consistent with a collinear spin density wave phase.  Interplane coupling is likely much weaker given the lattice structure making the wavelength of the SDW more susceptible to either entropic effects or subtle changes in the electronic structure along this direction.  Excluding GdRu$_2$Si$_2$ and GdRu$_2$Ge$_2$, a number of previously studied tetragonal Gd-based intermetallics are known to host antiferromagetic ground states where the moments align in a basal plane, such as Gd$_2$IrIn$_8$~\cite{Granado2004} or Gd$_2$CoGa$_8$ GdRhIn$_5$~\cite{Granado2006} with examples of out-of-plane moment alignment in the layered compound GdMn$_2$Ge$_2$~\cite{Granovsky2010}. A similar spin density wave state was found in GdNi$_2$B$_2$C~\cite{Detlefs1996} with an initial modulation that evolves upon cooling that is eventually interrupted by a spin-reorientation transition. A change in the moment direction of GdNiSn$_4$ is not resolved in our current data below $T_{AF1}$, though detecting a small, partial projection away from the $a$-axis may beneath the resolution of our measurements.

A similar square lattice lanthanide compound EuAl$_4$ displays a cascade of magnetic phases in temperature and field~\cite{Kaneko2021}, which include both non-collinear and SDW orders at zero field~\cite{Vibhakar2024} in addition to a charge density wave (CDW) state~\cite{Ramakrishnan2022}. While the isostructural system EuGa$_4$ is a simple antiferromagnet with no CDW phase~\cite{Kawasaki2016}, the lattice may be alloyed between the two extremes with Al substitution for Ga~\cite{Stavinoha2018}. Along this alloy line, EuGa$_2$Al$_2$ shows a CDW with a SDW that condenses into a spin cycloid at low temperature~\cite{Moya2022,Vibhakar2023}, and EuGa$_{2.4}$Al$_{1.6}$ has a ground state SDW with coexisting helices at low temperatures, but without a CDW~\cite{Littlehales2024}. As a result, it is hypothesized that a CDW distortion is required to stabilize the skyrmion phase in this system, and, indeed, a rectangular skyrmion lattice appears only in the orthorhombic phase of EuAl$_4$~\cite{Hayami2022a,Gen2023}. This highlights the role of lattice symmetry in the formation of magnetic phases. However, the corresponding tetragonal-to-orthorhombic distortion in EuAl$_4$ is much more subtle than the monoclinic distortion endemic to the lattice of GdNiSn$_4$. This suggests that the two-fold in-plane rotational symmetry breaking is too strong in the case of GdNiSn$_4$ and serves to pin a single in-plane propagation wave vector.

In summary, we have investigated the zero-field magnetic structure of the newly reported square-net compound GdNiSn$_4$ \cite{Zhang2025}.  REXS data reveal a collinear spin density wave state that is modulated both along the short in-plane and long interplane directions and whose moments are predominantly aligned along the in-plane propagation wave vector.  The wave length of this longitudinal SDW slides upon cooling until an intermediate, first-order magnetic transition is reached and the out-of-plane period of modulation is shortened.  Both magnetic transitions couple to electron transport in this compound and are likely driven via Fermi surface nesting effects.  Our work provides an important zero-field picture of magnetic symmetry breaking in this compound essential for future work modeling field-dependent effects.

\section{Acknowledgments}
This work was supported via AFOSR award FA9550-23-1-0635. We acknowledge beamtime at Diamond light source under proposal ID MM39446.

\bibliography{GdNiSn4_REXS}

@Article{Muehlbauer2009,
  author    = {M{\"u}hlbauer, S. and Binz, B. and Jonietz, F. and Pfleiderer, C. and Rosch, A. and Neubauer, A. and Georgii, R. and B{\"o}ni, P.},
  journal   = {Science},
  title     = {Skyrmion Lattice in a Chiral Magnet},
  year      = {2009},
  issn      = {1095-9203},
  month     = feb,
  number    = {5916},
  pages     = {915--919},
  volume    = {323},
  doi       = {10.1126/science.1166767},
  publisher = {American Association for the Advancement of Science (AAAS)},
}

@Article{Yu2010,
  author    = {Yu, X. Z. and Onose, Y. and Kanazawa, N. and Park, J. H. and Han, J. H. and Matsui, Y. and Nagaosa, N. and Tokura, Y.},
  journal   = {Nature},
  title     = {Real-space observation of a two-dimensional skyrmion crystal},
  year      = {2010},
  issn      = {1476-4687},
  month     = jun,
  number    = {7300},
  pages     = {901--904},
  volume    = {465},
  doi       = {10.1038/nature09124},
  publisher = {Springer Science and Business Media LLC},
}

@article{tokura2020magnetic,
  title={Magnetic skyrmion materials},
  author={Tokura, Yoshinori and Kanazawa, Naoya},
  journal={Chemical Reviews},
  volume={121},
  number={5},
  pages={2857--2897},
  year={2020},
  publisher={ACS Publications}
}

@misc{Supplemental,
	title = {Supplemental Information},
	url = {Can be found here},
	
}

@Article{Yoshimochi2024,
  author    = {Yoshimochi, H. and Takagi, R. and Ju, J. and Khanh, N. D. and Saito, H. and Sagayama, H. and Nakao, H. and Itoh, S. and Tokura, Y. and Arima, T. and Hayami, S. and Nakajima, T. and Seki, S.},
  journal   = {Nature Physics},
  title     = {Multistep topological transitions among meron and skyrmion crystals in a centrosymmetric magnet},
  year      = {2024},
  issn      = {1745-2481},
  month     = apr,
  number    = {6},
  pages     = {1001--1008},
  volume    = {20},
  doi       = {10.1038/s41567-024-02445-9},
  publisher = {Springer Science and Business Media LLC},
}

@Article{Moya2022,
  author    = {Moya, Jaime M. and Lei, Shiming and Clements, Eleanor M. and Kengle, Caitlin S. and Sun, Stella and Allen, Kevin and Li, Qizhi and Peng, Y. Y. and Husain, Ali A. and Mitrano, Matteo and Krogstad, Matthew J. and Osborn, Raymond and Puthirath, Anand B. and Chi, Songxue and Debeer-Schmitt, L. and Gaudet, J. and Abbamonte, P. and Lynn, Jeffrey W. and Morosan, E.},
  journal   = {Physical Review Materials},
  title     = {Incommensurate magnetic orders and topological {Hall} effect in the square-net centrosymmetric {EuGa}$_2${Al}$_2$ system},
  year      = {2022},
  issn      = {2475-9953},
  month     = jul,
  number    = {7},
  pages     = {074201},
  volume    = {6},
  doi       = {10.1103/physrevmaterials.6.074201},
  publisher = {American Physical Society (APS)},
}

@Article{Kurumaji2019,
  author    = {Kurumaji, Takashi and Nakajima, Taro and Hirschberger, Max and Kikkawa, Akiko and Yamasaki, Yuichi and Sagayama, Hajime and Nakao, Hironori and Taguchi, Yasujiro and Arima, Taka-hisa and Tokura, Yoshinori},
  journal   = {Science},
  title     = {Skyrmion lattice with a giant topological {Hall} effect in a frustrated triangular-lattice magnet},
  year      = {2019},
  issn      = {1095-9203},
  month     = aug,
  number    = {6456},
  pages     = {914--918},
  volume    = {365},
  doi       = {10.1126/science.aau0968},
  publisher = {American Association for the Advancement of Science (AAAS)},
}

@Article{Hirschberger2019,
  author    = {Hirschberger, Max and Nakajima, Taro and Gao, Shang and Peng, Licong and Kikkawa, Akiko and Kurumaji, Takashi and Kriener, Markus and Yamasaki, Yuichi and Sagayama, Hajime and Nakao, Hironori and Ohishi, Kazuki and Kakurai, Kazuhisa and Taguchi, Yasujiro and Yu, Xiuzhen and Arima, Taka-hisa and Tokura, Yoshinori},
  journal   = {Nature Communications},
  title     = {Skyrmion phase and competing magnetic orders on a breathing kagome lattice},
  year      = {2019},
  issn      = {2041-1723},
  month     = dec,
  number    = {1},
  volume    = {10},
  doi       = {10.1038/s41467-019-13675-4},
  publisher = {Springer Science and Business Media LLC},
}

@Article{Kaneko2021,
  author    = {Kaneko, Koji and Kawasaki, Takuro and Nakamura, Ai and Munakata, Koji and Nakao, Akiko and Hanashima, Takayasu and Kiyanagi, Ryoji and Ohhara, Takashi and Hedo, Masato and Nakama, Takao and \={O}nuki, Yoshichika},
  journal   = {Journal of the Physical Society of Japan},
  title     = {Charge-Density-Wave Order and Multiple Magnetic Transitions in Divalent Europium Compound {EuAl}$_4$},
  year      = {2021},
  issn      = {1347-4073},
  month     = jun,
  number    = {6},
  pages     = {064704},
  volume    = {90},
  doi       = {10.7566/jpsj.90.064704},
  publisher = {Physical Society of Japan},
}

@Article{Khanh2022,
  author    = {Khanh, Nguyen Duy and Nakajima, Taro and Hayami, Satoru and Gao, Shang and Yamasaki, Yuichi and Sagayama, Hajime and Nakao, Hironori and Takagi, Rina and Motome, Yukitoshi and Tokura, Yoshinori and Arima, Taka‐hisa and Seki, Shinichiro},
  journal   = {Advanced Science},
  title     = {Zoology of Multiple‐${Q}$ Spin Textures in a Centrosymmetric Tetragonal Magnet with Itinerant Electrons},
  year      = {2022},
  issn      = {2198-3844},
  month     = jan,
  number    = {10},
  volume    = {9},
  doi       = {10.1002/advs.202105452},
  publisher = {Wiley},
}

@Misc{Zhang2025,
  author = {Zhang, Xin and Lee, Scott B. and Chatterjee, Sudipta and Pi, Hanqi and Yang, Yi and Katmer, Fatmag\"{u}l and Ward, Emily G. and Widdowson, Daniel E. and Tam,  Charles C. and Schwarz,  Sarah and Pollak, Connor J. and Moya, Jaime M. and  Skorupskii, Grigorii and Kurlin, Vitaliy A. and Wilson, Stephen D. and  
Bernevig, B. Andrei and Schoop, Leslie M.},
  title  = {Identification of an Unreported Structure Type in {GdNiSn}$_4$ and Its Implications for Materials Prediction},
  year   = {2026},
  month     = mar,
  archivePrefix = {arXiv},
  eprint = {2603.05613},
  primaryClass = {cond-mat.mtrl-sci},
}

@Article{Hill1996,
  author    = {Hill, J. P. and McMorrow, D. F.},
  journal   = {Acta Crystallographica Section A Foundations of Crystallography},
  title     = {X-ray Resonant Exchange Scattering: Polariztaion Dependence and Correlation Function},
  year      = {1996},
  issn      = {0108-7673},
  month     = mar,
  number    = {2},
  pages     = {236--244},
  volume    = {52},
  doi       = {10.1107/s0108767395012670},
  publisher = {International Union of Crystallography (IUCr)},
}

@Article{Vibhakar2024,
  author    = {Vibhakar, Anuradha M. and Khalyavin, Dmitry D. and Orlandi, Fabio and Moya, Jamie M. and Lei, Shiming and Morosan, Emilia and Bombardi, Alessandro},
  journal   = {Communications Physics},
  title     = {Spontaneous reversal of spin chirality and competing phases in the topological magnet {EuAl}$_4$},
  year      = {2024},
  issn      = {2399-3650},
  month     = sep,
  number    = {1},
  volume    = {7},
  doi       = {10.1038/s42005-024-01802-7},
  publisher = {Springer Science and Business Media LLC},
}

@Article{Hayami2022a,
  author    = {Hayami, Satoru},
  journal   = {Journal of Physics: Materials},
  title     = {Orthorhombic distortion and rectangular skyrmion crystal in a centrosymmetric tetragonal host},
  year      = {2022},
  issn      = {2515-7639},
  month     = dec,
  number    = {1},
  pages     = {014006},
  volume    = {6},
  doi       = {10.1088/2515-7639/acab89},
  publisher = {IOP Publishing},
}

@Article{Porter2023,
  author    = {Porter, Zach and Pokharel, Ganesh and Kim, Jong-Woo and Ryan, Phillip J. and Wilson, Stephen D.},
  journal   = {Physical Review B},
  title     = {Incommensurate magnetic order in the ${Z}_2$ kagome metal {GdV}$_6${Sn}$_6$},
  year      = {2023},
  issn      = {2469-9969},
  month     = jul,
  number    = {3},
  pages     = {035134},
  volume    = {108},
  doi       = {10.1103/physrevb.108.035134},
  publisher = {American Physical Society (APS)},
}

@Article{Kawasaki2016,
  author    = {Kawasaki, Takuro and Kaneko, Koji and Nakamura, Ai and Aso, Naofumi and Hedo, Masato and Nakama, Takao and Ohhara, Takashi and Kiyanagi, Ryoji and Oikawa, Kenichi and Tamura, Itaru and Nakao, Akiko and Munakata, Koji and Hanashima, Takayasu and \={O}nuki, Yoshichika},
  journal   = {Journal of the Physical Society of Japan},
  title     = {Magnetic Structure of Divalent Europium Compound {EuGa}$_4$ Studied by Single-Crystal Time-of-Flight Neutron Diffraction},
  year      = {2016},
  issn      = {1347-4073},
  month     = nov,
  number    = {11},
  pages     = {114711},
  volume    = {85},
  doi       = {10.7566/jpsj.85.114711},
  publisher = {Physical Society of Japan},
}

@Article{Stavinoha2018,
  author    = {Stavinoha, Macy and Cooley, Joya A. and Minasian, Stefan G. and McQueen, Tyrel M. and Kauzlarich, Susan M. and Huang, C.-L. and Morosan, E.},
  journal   = {Physical Review B},
  title     = {Charge density wave behavior and order-disorder in the antiferromagnetic metallic series {Eu(Ga}$_{1-x}${Al}$_x$)$_4$},
  year      = {2018},
  issn      = {2469-9969},
  month     = may,
  number    = {19},
  pages     = {195146},
  volume    = {97},
  doi       = {10.1103/physrevb.97.195146},
  publisher = {American Physical Society (APS)},
}

@Article{Vibhakar2023,
  author    = {Vibhakar, A. M. and Khalyavin, D. D. and Moya, J. M. and Manuel, P. and Orlandi, F. and Lei, S. and Morosan, E. and Bombardi, A.},
  journal   = {Physical Review B},
  title     = {Competing charge and magnetic order in the candidate centrosymmetric skyrmion host {EuGa}$_2${Al}$_2$},
  year      = {2023},
  issn      = {2469-9969},
  month     = sep,
  number    = {10},
  pages     = {l100404},
  volume    = {108},
  doi       = {10.1103/physrevb.108.l100404},
  publisher = {American Physical Society (APS)},
}

@Article{Littlehales2024,
  author    = {Littlehales, M. T. and Moody, S. H. and Bereciartua, P. J. and Mayoh, D. A. and Parkin, Z. B. and Blundell, T. J. and Unsworth, E. and Francoual, S. and Balakrishnan, G. and Venero, D. Alba and Hatton, P. D.},
  journal   = {Physical Review Research},
  title     = {Spin density waves and ground state helices in {EuGa}$_{2.4}${Al}$_{1.6}$},
  year      = {2024},
  issn      = {2643-1564},
  month     = jul,
  number    = {3},
  pages     = {l032015},
  volume    = {6},
  doi       = {10.1103/physrevresearch.6.l032015},
  publisher = {American Physical Society (APS)},
}

@Article{Ramakrishnan2022,
  author    = {Ramakrishnan, Sitaram and Kotla, Surya Rohith and Rekis, Toms and Bao, Jin-Ke and Eisele, Claudio and Noohinejad, Leila and Tolkiehn, Martin and Paulmann, Carsten and Singh, Birender and Verma, Rahul and Bag, Biplab and Kulkarni, Ruta and Thamizhavel, Arumugam and Singh, Bahadur and Ramakrishnan, Srinivasan and van Smaalen, Sander},
  journal   = {IUCrJ},
  title     = {Orthorhombic charge density wave on the tetragonal lattice of {EuAl}$_4$},
  year      = {2022},
  issn      = {2052-2525},
  month     = apr,
  number    = {3},
  pages     = {378--385},
  volume    = {9},
  doi       = {10.1107/s2052252522003888},
  publisher = {International Union of Crystallography (IUCr)},
}

@Article{Hirschberger2020,
  author    = {Hirschberger, Max and Spitz, Leonie and Nomoto, Takuya and Kurumaji, Takashi and Gao, Shang and Masell, Jan and Nakajima, Taro and Kikkawa, Akiko and Yamasaki, Yuichi and Sagayama, Hajime and Nakao, Hironori and Taguchi, Yasujiro and Arita, Ryotaro and Arima, Taka-hisa and Tokura, Yoshinori},
  journal   = {Physical Review Letters},
  title     = {Topological {Nernst} Effect of the Two-Dimensional Skyrmion Lattice},
  year      = {2020},
  issn      = {1079-7114},
  month     = aug,
  number    = {7},
  pages     = {076602},
  volume    = {125},
  doi       = {10.1103/physrevlett.125.076602},
  publisher = {American Physical Society (APS)},
}

@Article{Yasui2020,
  author    = {Yasui, Yuuki and Butler, Christopher J. and Khanh, Nguyen Duy and Hayami, Satoru and Nomoto, Takuya and Hanaguri, Tetsuo and Motome, Yukitoshi and Arita, Ryotaro and Arima, Taka-hisa and Tokura, Yoshinori and Seki, Shinichiro},
  journal   = {Nature Communications},
  title     = {Imaging the coupling between itinerant electrons and localised moments in the centrosymmetric skyrmion magnet {GdRu}$_2${Si}$_2$},
  year      = {2020},
  issn      = {2041-1723},
  month     = nov,
  number    = {1},
  volume    = {11},
  doi       = {10.1038/s41467-020-19751-4},
  publisher = {Springer Science and Business Media LLC},
}

@InProceedings{Collins2010,
  author    = {Collins, S. P. and Bombardi, A. and Marshall, A. R. and Williams, J. H. and Barlow, G. and Day, A. G. and Pearson, M. R. and Woolliscroft, R. J. and Walton, R. D. and Beutier, G. and Nisbet, G. and Garrett, R. and Gentle, I. and Nugent, K. and Wilkins, S.},
  booktitle = {AIP Conference Proceedings},
  title     = {Diamond Beamline {I16} (Materials \& Magnetism)},
  year      = {2010},
  pages     = {303--306},
  publisher = {AIP},
  doi       = {10.1063/1.3463196},
  issn      = {0094-243X},
}

@Article{Garnier1996,
  author    = {Garnier, A. and Gignoux, D. and Schmitt, D. and Shigeoka, T.},
  journal   = {Physica B: Condensed Matter},
  title     = {Giant magnetic anisotropy in tetragonal {GdRu}$_2${Ge}$_2$ and {GdRu}$_2${Si}$_2$},
  year      = {1996},
  issn      = {0921-4526},
  month     = may,
  number    = {1–3},
  pages     = {80--86},
  volume    = {222},
  doi       = {10.1016/0921-4526(96)00010-5},
  publisher = {Elsevier BV},
}

@Article{Garnier1995,
  author    = {Garnier, A. and Gignoux, D. and Iwata, N. and Schmitt, D. and Shigeoka, T. and Zhang, F.Y.},
  journal   = {Journal of Magnetism and Magnetic Materials},
  title     = {Anisotropic metamagnetism in {GdRu}$_2${Si}$_2$},
  year      = {1995},
  issn      = {0304-8853},
  month     = feb,
  pages     = {899--900},
  volume    = {140–144},
  doi       = {10.1016/0304-8853(94)00783-7},
  publisher = {Elsevier BV},
}

@Article{Gen2023,
  author    = {Gen, Masaki and Takagi, Rina and Watanabe, Yoshito and Kitou, Shunsuke and Sagayama, Hajime and Matsuyama, Naofumi and Kohama, Yoshimitsu and Ikeda, Akihiko and Ōnuki, Yoshichika and Kurumaji, Takashi and Arima, Taka-hisa and Seki, Shinichiro},
  journal   = {Physical Review B},
  title     = {Rhombic skyrmion lattice coupled with orthorhombic structural distortion in {EuAl}$_4$},
  year      = {2023},
  issn      = {2469-9969},
  month     = jan,
  number    = {2},
  pages     = {l020410},
  volume    = {107},
  doi       = {10.1103/physrevb.107.l020410},
  publisher = {American Physical Society (APS)},
}

@Article{Khanh2020,
  author    = {Khanh, Nguyen Duy and Nakajima, Taro and Yu, Xiuzhen and Gao, Shang and Shibata, Kiyou and Hirschberger, Max and Yamasaki, Yuichi and Sagayama, Hajime and Nakao, Hironori and Peng, Licong and Nakajima, Kiyomi and Takagi, Rina and Arima, Taka-hisa and Tokura, Yoshinori and Seki, Shinichiro},
  journal   = {Nature Nanotechnology},
  title     = {Nanometric square skyrmion lattice in a centrosymmetric tetragonal magnet},
  year      = {2020},
  issn      = {1748-3395},
  month     = may,
  number    = {6},
  pages     = {444--449},
  volume    = {15},
  doi       = {10.1038/s41565-020-0684-7},
  publisher = {Springer Science and Business Media LLC},
}

@Article{Nakamura2015,
  author    = {Nakamura, Ai and Uejo, Taro and Honda, Fuminori and Takeuchi, Tetsuya and Harima, Hisatomo and Yamamoto, Etsuji and Haga, Yoshinori and Matsubayashi, Kazuyuki and Uwatoko, Yoshiya and Hedo, Masato and Nakama, Takao and \={O}nuki, Yoshichika},
  journal   = {Journal of the Physical Society of Japan},
  title     = {Transport and Magnetic Properties of {EuAl}$_4$ and {EuGa}$_4$},
  year      = {2015},
  issn      = {1347-4073},
  month     = dec,
  number    = {12},
  pages     = {124711},
  volume    = {84},
  doi       = {10.7566/jpsj.84.124711},
  publisher = {Physical Society of Japan},
}

@Article{Granado2004,
  author    = {Granado, E. and Pagliuso, P. G. and Giles, C. and Lora-Serrano, R. and Yokaichiya, F. and Sarrao, J. L.},
  journal   = {Physical Review B},
  title     = {Magnetic structure and fluctuations of {Gd}$_2${IrIn}$_8$: A resonant x-ray diffraction study},
  year      = {2004},
  issn      = {1550-235X},
  month     = apr,
  number    = {14},
  pages     = {144411},
  volume    = {69},
  doi       = {10.1103/physrevb.69.144411},
  publisher = {American Physical Society (APS)},
}

@Article{Granado2006,
  author    = {Granado, E. and Uchoa, B. and Malachias, A. and Lora-Serrano, R. and Pagliuso, P. G. and Westfahl, H.},
  journal   = {Physical Review B},
  title     = {Magnetic structure and critical behavior of {GdRhIn}$_5$: Resonant x-ray diffraction and renormalization group analysis},
  year      = {2006},
  issn      = {1550-235X},
  month     = dec,
  number    = {21},
  pages     = {214428},
  volume    = {74},
  doi       = {10.1103/physrevb.74.214428},
  publisher = {American Physical Society (APS)},
}

@Article{Takagi2022,
  author    = {Takagi, Rina and Matsuyama, Naofumi and Ukleev, Victor and Yu, Le and White, Jonathan S. and Francoual, Sonia and Mardegan, José R. L. and Hayami, Satoru and Saito, Hiraku and Kaneko, Koji and Ohishi, Kazuki and \={O}nuki, Yoshichika and Arima, Taka-hisa and Tokura, Yoshinori and Nakajima, Taro and Seki, Shinichiro},
  journal   = {Nature Communications},
  title     = {Square and rhombic lattices of magnetic skyrmions in a centrosymmetric binary compound},
  year      = {2022},
  issn      = {2041-1723},
  month     = mar,
  number    = {1},
  volume    = {13},
  doi       = {10.1038/s41467-022-29131-9},
  publisher = {Springer Science and Business Media LLC},
}

@Article{Detlefs1996,
  author    = {Detlefs, C. and Goldman, A. I. and Stassis, C. and Canfield, P. C. and Cho, B. K. and Hill, J. P. and Gibbs, D.},
  journal   = {Physical Review B},
  title     = {Magnetic structure of {GdNi}$_2${B}$_2${C} by resonant and nonresonant x-ray scattering},
  year      = {1996},
  issn      = {1095-3795},
  month     = mar,
  number    = {10},
  pages     = {6355--6361},
  volume    = {53},
  doi       = {10.1103/physrevb.53.6355},
  publisher = {American Physical Society (APS)},
}

@Article{Granovsky2010,
  author    = {Granovsky, S A and Kreyssig, A and Doerr, M and Ritter, C and Dudzik, E and Feyerherm, R and Canfield, P C and Loewenhaupt, M},
  journal   = {Journal of Physics: Condensed Matter},
  title     = {The magnetic order of {GdMn}$_2${Ge}$_2$ studied by neutron diffraction and x-ray resonant magnetic scattering},
  year      = {2010},
  issn      = {1361-648X},
  month     = may,
  number    = {22},
  pages     = {226005},
  volume    = {22},
  doi       = {10.1088/0953-8984/22/22/226005},
  publisher = {IOP Publishing},
}

@Article{Ozawa2017,
  author    = {Ozawa, Ryo and Hayami, Satoru and Motome, Yukitoshi},
  journal   = {Physical Review Letters},
  title     = {Zero-Field Skyrmions with a High Topological Number in Itinerant Magnets},
  year      = {2017},
  issn      = {1079-7114},
  month     = apr,
  number    = {14},
  pages     = {147205},
  volume    = {118},
  doi       = {10.1103/physrevlett.118.147205},
  publisher = {American Physical Society (APS)},
}

@Article{Wang2020,
  author    = {Wang, Zhentao and Su, Ying and Lin, Shi-Zeng and Batista, Cristian D.},
  journal   = {Physical Review Letters},
  title     = {Skyrmion Crystal from {RKKY} Interaction Mediated by {2D} Electron Gas},
  year      = {2020},
  issn      = {1079-7114},
  month     = may,
  number    = {20},
  pages     = {207201},
  volume    = {124},
  doi       = {10.1103/physrevlett.124.207201},
  publisher = {American Physical Society (APS)},
}

@Article{Hayami2017,
  author    = {Hayami, Satoru and Ozawa, Ryo and Motome, Yukitoshi},
  journal   = {Physical Review B},
  title     = {Effective bilinear-biquadratic model for noncoplanar ordering in itinerant magnets},
  year      = {2017},
  issn      = {2469-9969},
  month     = jun,
  number    = {22},
  pages     = {224424},
  volume    = {95},
  doi       = {10.1103/physrevb.95.224424},
  publisher = {American Physical Society (APS)},
}

\end{document}